\documentclass[twocolumn,showpacs,footnoteinbib]{revtex4}
\usepackage{amsmath,amssymb,epic,eepic}
\usepackage[dvips]{graphicx}
\usepackage{epsfig}
\usepackage{color}
\usepackage{graphicx}

\begin{document}

\title{Decoherence and relaxation of single electron excitations
in quantum Hall edge channels.}


\author{P.~Degiovanni$^{1}$}
\author{Ch. Grenier$^{1}$}
\author{G.~F\`eve$^{2,3}$}

\affiliation{(1) Universit\'e de Lyon, F\'ed\'eration de Physique Andr\'e Marie Amp\`ere,
CNRS UMR 5672 - Laboratoire de Physique de l'Ecole Normale Sup\'erieure de Lyon,
46 All\'ee d'Italie, 69364 Lyon Cedex 07,
France}

\affiliation{(2) Laboratoire Pierre Aigrain, Ecole Normale Sup\'erieure, 24 rue Lhomond, 75231 Paris Cedex
05, France}

\affiliation{(3) Laboratoire associ\'e aux universit\'es Pierre \& Marie Curie et Denis Diderot, CNRS - UMR 8551, Paris, France}

\begin{abstract}
A unified approach to decoherence and relaxation of energy resolved single electron excitations
in Integer Quantum Hall edge channels is presented. Within the bosonization framework, relaxation and
decoherence induced by interactions and capacitive coupling to an external linear circuit are
computed. An explicit connexion with high frequency transport properties
of a two terminal device formed
by the edge channel on one side and the linear circuit on the other side is established.
\end{abstract}

\pacs{73.23.-b,03.65.Yz,73.43.Cd,73.43.Lp}

\maketitle

The recent demonstrations of Mach-Zehnder interferometers (MZI)~\cite{Ji:2003-1,Neder:2007-2,Roulleau:2008-2} 
in the Integer Quantum Hall regime has
best illustrated the ballistic and phase coherent character of electronic propagation along the chiral quantum Hall 
edges over tens of microns. Recently, the
development of an on demand single electron source
\cite{Feve:2007-1} has opened the way to fundamental {\em electron quantum optics}
experiments involving single charge excitations such as the electronic Hanbury-Brown and Twiss
experiment \cite{Hanbury:1956-1}
or the Hong-Ou-Mandel experiment \cite{Hong:1987-1,Olkhovskaya:2008-1} in which
two indistinguishable electrons collide on a beam-splitter.
But contrary to photons, electrons interact with their
electromagnetic environment and with other electrons present the
Fermi sea. This results in relaxation and decoherence of single
electron excitations above the ground state that deeply questions the
whole electron quantum optics concept. This has
been emphasized in MZI where,
despite insensitivity to electron-source time
coherence, decoherence along the chiral edges drastically reduces
the contrast of interferences
\cite{Seelig:2001-1,Chalker:2007-1,Roulleau:2008-1,Sukhorukov:2007-2,Levkivskyi:2008-1,Neuenhahn:2009-1}.

Recently, energy resolved electronic detection using quantum dots
has also been demonstrated \cite{Altimiras:2009-1}, thus opening the
way to energy relaxation studies in quantum Hall edge channels. Combining such
measurements with the on demand single electron source will enable experimental testing
of  the basic problem of quasi-particle relaxation originally considered by Laudau 
in Fermi liquid theory~\cite{Nozieres-Pines}. 

In this Rapid Communication, we present a full many body solution to this problem  in quantum Hall edge channels. To
address the above basic issues, we have developed
a unified approach of high frequency transport and decoherence and relaxation of 
coherent single electron excitations showing the central role of plasmon scattering. 
This approach opens the way to an in depth understanding of the nature of quasiparticles
in various Integer Quantum Hall edge channels and provides an efficient 
theoretical framework for electron quantum optics. To illustrate our formalism, we clarify the 
role of the electronic Fermi sea in single electron decoherence, explicitely showing the 
interpolation from a Pauli blockade regime at low energy to a dynamical 
Coulomb blockade like regime at high energy in which the Fermi sea
plays the role of an effective environment. 

\paragraph*{Model}

\begin{figure}
\includegraphics[angle=270,width=8cm]{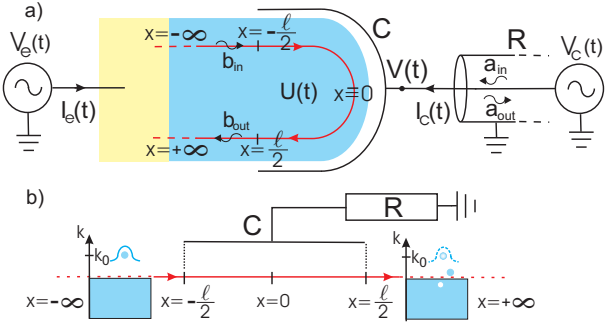}
\caption{\label{fig:schema} (Color online) (a) Two terminal device built from an
edge channel coupled to an RC circuit. In the interaction area
$|x|\leq  l/2$, the edge channel (in red) at internal potential
$U(t)$ is capacitively coupled to the gate of capacitance $C$ at
potential $V(t)$.  (b) A single electron is injected in the
$x<-l/2$ region and propagates along the edge channel.
Interactions within $|x|\leq l/2$ create electron/hole pairs
within the edge channel and photons in the RC circuit leading to
decoherence.}
\end{figure}

Let us consider the specific example of a chiral edge channel
capacitively coupled to an external gate of size $l$ connected to a
resistance $R$ representing a dissipative external circuit (see Fig.
\ref{fig:schema}). In the integer quantum Hall
regime, bosonization expresses the electron creation operator at
point $x$ along the edge, $\psi^{\dagger}(x)$, in terms of a chiral
bosonic field $\phi(x)$ as:
\begin{equation}\label{eq:chiral:bosonization}
\psi^\dagger(x)=\frac{U^\dagger}{\sqrt{2\pi
a}}e^{-i\sqrt{4\pi}\phi(x)},
\end{equation}
where $a$ is a short distance cutoff and $U^\dagger$ ensures
fermionic anticommutation relations. The bosonic field determines
the electron density $n(x)=(\partial_{x}\phi)(x)/\sqrt{\pi}$. In the
presence of an external voltage $U(x,t)$ along the edge, its
equation of motion is:
\begin{equation}
\label{eq:chiral:eqsmotion}
\left(\partial_{t}+v_{F}\partial_{x}\right)\phi(x,t)=\frac{e\sqrt{\pi}}{h}\,U(x,t)\,.
\end{equation}
where $v_F$ is the electron Fermi velocity along the edge. Before
and after the interaction region $|x|\leq l/2$, this chiral field
propagates freely and thus defines input ($j=\mathrm{in}$ for
$x\leq -l/2$) and output ($j=\mathrm{out}$ for $x\geq l/2$)
plasmon modes :
\begin{equation}
\label{eq:b-modes} \phi_{j}(x,t)=\frac{-i}{\sqrt{4\pi}}
\int_{0}^\infty \frac{d\omega}{\sqrt{\omega}} ( b_{j}(\omega)
e^{i \omega (x /v_F - t)}-\mathrm{h.c} )\,.
\end{equation}
In the interaction area $|x|\leq  l/2$,  the edge is capacitively
coupled to the gate forming a capacitor of capacitance $C$. Following B\"{u}ttiker \textit{et al.} \cite{Pretre:1996-1}, we assume the potential within the edge to
be uniform \footnote{This hypothesis is valid for $\omega/2\pi \lesssim v_{F}/l$ but the general framework presented here
is model independent and thus applies to more realistic models of circuit/edge channel coupling.} 
and denoted by $U(t)$ \cite{Pretre:1996-1}. The total charge stored within the interaction region
\begin{equation}
\label{eq:neutrality}
Q(t)=-e\int_{-l/2}^{l/2} n(x,t)\,dx
\end{equation}
is proportional to the voltage drop between
the gate at potential $V(t)$ and the edge channel:
\begin{equation}
\label{eq:voltage-drop}
C(U(t)-V(t))=Q(t)\,.
\end{equation}
Following \cite{Yurke:1984-1,Degio25}, the
resistance will be modeled as a quantum transmission line with
characteristic impedance $Z=R$. Its degrees of freedom are described
by input and output photon modes $a_{j}(\omega)$
($j=\mathrm{in}$ or $\mathrm{out}$) propagating along the line.
The total charge stored within the line is, by neutrality of the RC
circuit, equal to $Q(t)$ and expressed in terms of photon modes by:
\begin{equation}
\label{eq:a-modes:Q}
Q(t)=\sqrt{\frac{\hbar}{4\pi R}}\int_{0}^\infty
((a_{\mathrm{in}}+a_{\mathrm{out}})(\omega)e^{-i\omega t}
+\mathrm{h.c.})
\frac{d\omega}{\sqrt{\omega}}\,.
\end{equation}
The voltage of the gate $V(t)$ is also expressed as:
\begin{equation}
\label{eq:a-modes:V}
V(t)= eR\sqrt{\frac{R_{K}}{2R}}\int_{0}^\infty \sqrt{\omega}(i(a_{\mathrm{in}}-a_{\mathrm{out}})(\omega)
e^{-i\omega t}+\mathrm{h.c.})\frac{d\omega}{2\pi}
\end{equation}
Using \eqref{eq:voltage-drop}, \eqref{eq:a-modes:Q} and
\eqref{eq:a-modes:V}, the voltage $U(t)$ can be expressed in terms of the $a$ modes. Solving for the edge equation of motion \eqref{eq:chiral:eqsmotion} gives a first linear equation relating the input and output $a$ and $b$ modes. It also expresses
the chiral field and total charge within the interaction region in terms of these modes.
Using this in eq. \eqref{eq:neutrality} leads to a second linear equation between the input and output $a$ and $b$ modes.
This leads to the unitary plasmon to photon scattering matrix $S(\omega)= (S_{\alpha\beta}(\omega))$ ($\alpha$, $\beta$ equal to $c$ for the circuit or $e$ for the edge)
at frequency $\omega/2\pi$. In the following $t(\omega)=S_{ee}(\omega)$ 
will denote the edge plasmon transmission amplitude. As discussed in the case of 1D (non chiral) wires~\cite{Blanter:1998-1}, 
it determines the finite frequency admittance. 
Obviously, this quantity is model dependent but our discussion 
of its connection with single electron relaxation is valid in full generality.

\paragraph*{High frequency admittances} The admittance matrix at frequency $\omega$
relates the Fourier component $I_{\alpha}(\omega)$ of the
incoming currents through the various leads $\alpha$ to the voltages
$V_{\beta}(\omega)$ in the linear response regime:
\begin{equation}
\label{eq:admittances:definition}
I_{\alpha}(\omega)=\sum_{\beta}g_{\alpha
\beta}(\omega)\,V_{\beta}(\omega)\,.
\end{equation}
In the present case, the input modes can be related to the
voltages applied to the edge reservoir and to the RC circuit. First of all,
the applied time dependent voltage $V_{c}(t)$ creates an incoming
coherent state within the transmission line  caracterized by the average value of
the incoming $a$ modes:
\begin{equation}
\label{eq:a-modes:voltage-in}
\langle a_{\mathrm{in}}(\omega)\rangle = \frac{-i}{\sqrt{\omega}}\sqrt{\frac{R_{K}}{2R}}\frac{eV_{c}(\omega)}{h}\,.
\end{equation}
For the edge channel, modeling the reservoir as a uniform time
dependent voltage $V_{e}(t)$ applied from $x=-\infty$ to $x=-l/2$
shows that it injects in the $[-l/2,l/2]$ region a plasmon coherent
state fully characterized by the average value of the incoming $b$
modes \cite{Safi:1999-1}:
\begin{equation}
\label{eq:b-modes:voltage-input} \langle b_{\mathrm{in}}(\omega)\rangle =
-\frac{e}{h}\frac{V_{e}(\omega)}{\sqrt{\omega}} \,e^{i\omega l/2 v_F}\,.
\end{equation}
The finite frequency admittance can then be expressed in terms of
the plasmon to photon scattering matrix and reciprocally uniquely
determines it. In the present case, due to total screening of
charges in the edge channel by the gate, this $2\times 2$ admittance
matrix satisfies gauge invariance and charge conservation
\cite{Pretre:1996-1}: $g_{ee}=g_{cc}=-g_{ec}=-g_{ce}$ and is thus
determined by a unique admittance:
\begin{equation}
\label{eq:admittance:gee}
 g(\omega)=g_{ee}(\omega) =
\frac{e^2}{h}(1-t(\omega)\,e^{i\omega l/ v_F}),
\end{equation}
which is in turn entirely determined by the plasmon
transmission amplitude $t(\omega)$. At low frequency and up to
second order in $\omega$, the admittance is
equivalent to the serial addition of a electrochemical capacitance
$C_{\mu}$ and a charge relaxation resistance $R_q$. Here, $C_{\mu}$ is the series
addition of the quantum capacitance $C_{q}=l/v_{F}R_{K}$ (with $R_K = \frac{h}{e^2}$)
and the geometric capacitance \cite{Pretre:1996-1}. The charge relaxation resistance is
the sum of the circuit resistance and the half-quantum $h/2e^2$ \cite{Buttiker:1993-2,Gabelli:2006-1}
expected for a single mode conductor. 

\paragraph*{Electron relaxation and decoherence} Let us now consider the evolution of a single electron
initially prepared (see Fig.\,\ref{fig:schema}-b) in a coherent wave
packet. Starting from the zero
temperature 
Fermi sea $|F\rangle$ , the many body state with
one additional electron in the normalized wavepacket $\varphi(x)$ above the Fermi sea is given by
$|\varphi,F\rangle=\int \varphi(x)\psi^\dagger(x)|F\rangle\,dx$.

Single particle spatial coherence and density are respectively described by the off
diagonal and diagonal components of the single particle reduced density
operator $\mathcal{G}^{(e)}_{\rho}(x,y)=\mathrm{Tr}(\psi(x)\ldotp
\rho \ldotp \psi^\dagger(y))$ in which $\rho$ denotes the many body
electronic density operator. Going to Fourier space then shows that the single electron
momentum distribution is encoded in the diagonal of the single particle density operator in momentum space.

Before entering the interaction region, the initial many body
density operator is given by
$\rho_{i}=|\varphi,F\rangle\langle\varphi,F|$ and then
$\mathcal{G}^{(e)}_{\rho_{i}}(x,y)=\mathcal{G}^{(e)}_{F}(x,y) +
\varphi(y)^*\varphi(x)$ where the first contribution
$\mathcal{G}^{(e)}_{F}(x,y) =\int
n_{F}(k)e^{ik(x-y)}\frac{dk}{2\pi}$ comes from the Fermi sea ($n_F(k) = \langle c^{\dagger}_{k}c_{k}\rangle_{F}$) whereas the second one is associated with the coherent  single electron excitation.

Using \eqref{eq:chiral:bosonization}, the state
$|\varphi,F\rangle$ appears as a superposition of incoming plasmon
coherent states. Since the interaction region elastically scatters
plasmon to photons, the resulting outcoming external circuit + edge
quantum state is an entangled superposition of coherent plasmon and photon states.
Tracing out over the circuit's degrees of freedom leads to the exact
many body electron state after time $t$ such that the initial wave
packet has flown through the interacting region:
\begin{eqnarray}
\label{eq:rhof}
\rho_{f} & = & \int dy_{+}dy_{-} \varphi(y_{+})\varphi^*(y_{-})
\mathcal{D}_{\mathrm{ext}}(y_+-y_-) \nonumber\\
& \times & \widetilde{\psi}^\dagger(y_{+}+v_{F}t)
|F\rangle
\langle F|\widetilde{\psi}(y_{-}+v_{F}t)
\end{eqnarray}
where $\mathcal{D}_{\mathrm{ext}}$ is the extrinsic decoherence associated with photon
emission into the external circuit 
\begin{equation}
\label{eq:decoherence:extrinsic}
\mathcal{D}_{\mathrm{ext}}(\Delta y) = \exp{\left(\int_{0}^{+\infty}|r(\omega)|^2\,(e^{-i\frac{\omega\,\Delta y}{v_{F}}}-1)\,
\frac{d\omega}{\omega}\right)}
\end{equation}
where $r(\omega)=S_{ce}(\omega)$ denotes the plasmon to photon scattering amplitude.
In \eqref{eq:rhof}, $\widetilde{\psi}^\dagger(y)$ is a modified field operator:
\begin{equation}
\widetilde{\psi}^\dagger(y)=\frac{U^\dagger}{\sqrt{2\pi a}}\exp{\left(
\int_0^\infty (t(\omega)b(\omega)\,e^{i\omega y/v_F} - \mathrm{h.c.})\frac{d\omega}{\sqrt{\omega}}
\right)}\, .
\end{equation}
When $t(\omega)$ is a pure phase linear in
$\omega$, $\widetilde{\psi}^\dagger (y)$ is a spatially translated fermion field operator.
Any other $\omega$ dependence of $t(\omega)$ leads to the creation of additional 
electron/hole pairs that cannot be absorbed in 
a simple translation.

\begin{figure}
\includegraphics[angle=270,width=8.5cm]{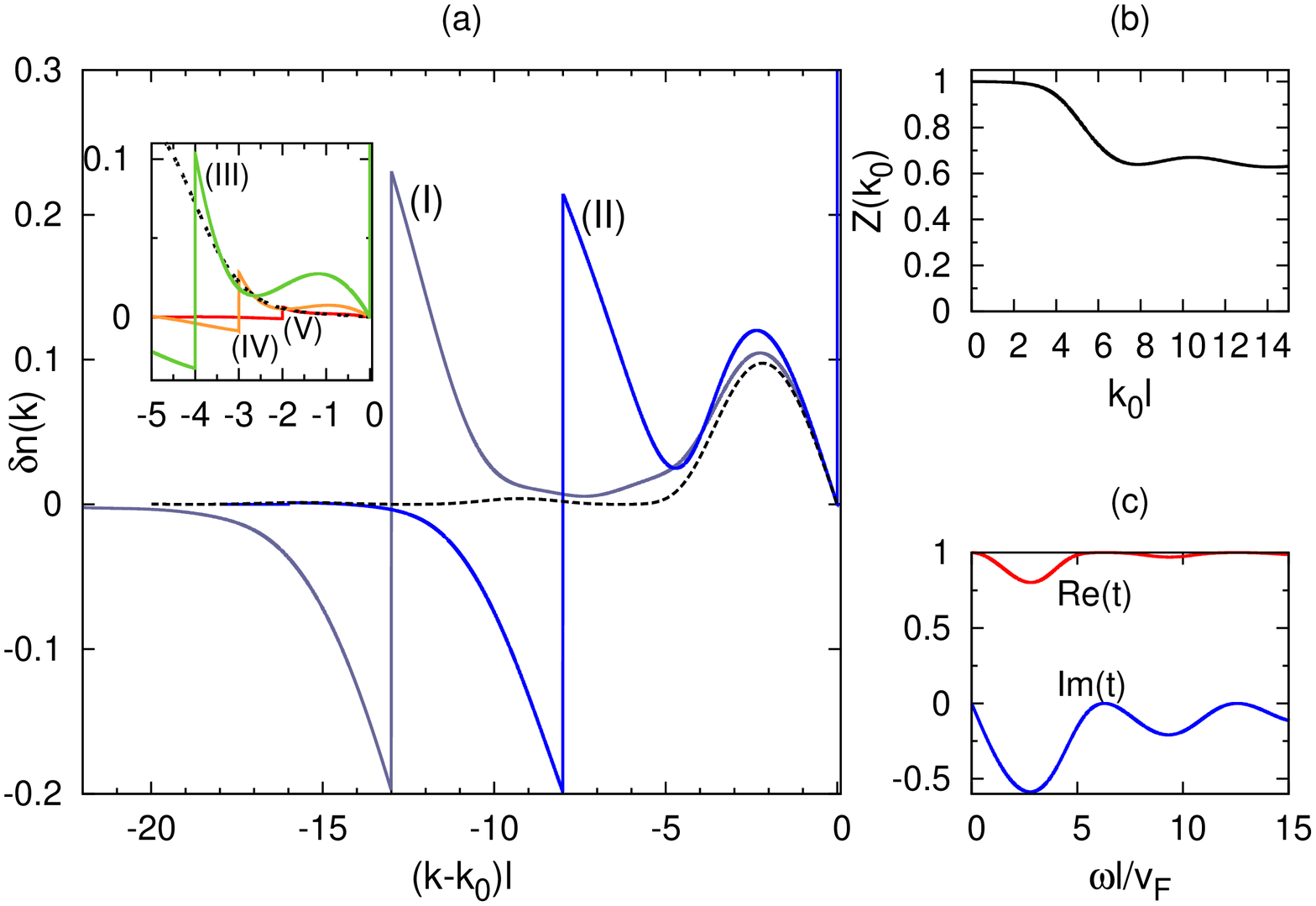}
\caption{\label{fig:Relaxation1} (Color online) (a) Single electron energy relaxation for $R=50\ \Omega$ and $l/v_{F}R_{K}C = 0.5$ plotted against $(k-k_0)l$ for (I) $k_0l = 13$ and (II) $k_0l = 8$.
The dashed curve corresponds to the single particle result \eqref{eq:single particle UV} valid at high energies. 
Inset shows small values of $k_{0}l$: (III) $4$, (IV) $3$ and (V) $2$. The dotted line corresponds to the simple relaxation model  $\delta n_{k_{0}}^r(k) = -Z{'}(k_0-k)$. (b) Quasiparticle peak $Z(k_0)$. (c) Real and imaginary part of
the plasmon transmission amplitude $t(\omega)$.}
\end{figure}

\begin{figure}
\includegraphics[angle=270,width=8.5cm]{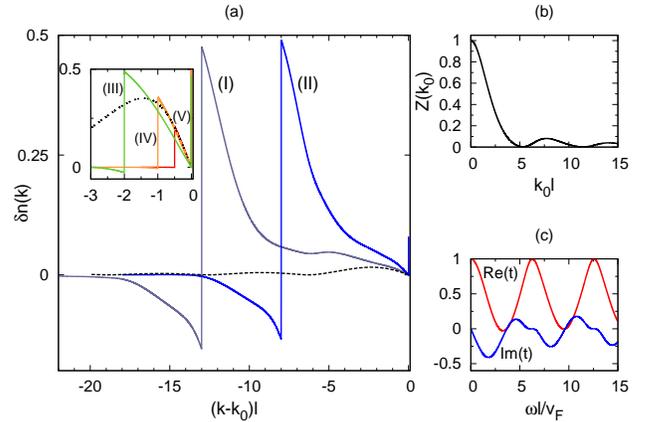}
\caption{\label{fig:Relaxation2}  (Color online) (a) Single electron energy relaxation for $R/R_K=0.5$ and $l/v_{F}R_{K}C = 0.5$ plotted against $(k-k_0)l$ for (I) $k_0l = 13$ and (II) $k_0l = 8$. The dashed curve corresponds to the single particle result \eqref{eq:single particle UV} valid at high energies. Inset shows small values of $k_{0}l$: (III) $2$, (IV) $1$ and (V) $0.5$. The dotted line corresponds to the simple relaxation model  $\delta n_{k_{0}}^r(k) = -Z{'}(k_0-k)$. (b) Quasiparticle peak $Z(k_0)$. (c) Real and imaginary part of
the plasmon transmission amplitude $t(\omega)$.}
\end{figure}

Relaxation of a single electron can then be discussed  by
considering an incident energy resolved wave function:
$\varphi_{k_{0}}(x)=e^{ik_{0}x}$. 
The complete momentum distribution consists in the Fermi sea contribution and a change $\delta n_{k_0}(k)$
that can be obtained from single particle coherence $\mathcal{G}^{(e)}_{k_0}$
after interaction with the RC-circuit:
\begin{equation}\label{eq:def delta n}
\int_{\mathbb{R}} e^{-ik(x-y)} \mathcal{G}^{(e)}_{k_0}(x,y)\,  d(x-y) =Ln_F(k)+\delta n_{k_{0}}(k)
\end{equation}
where $L$ is the total size of the system. The l.h.s. can be computed using bosonization thus leading to
explicit fully non perturbative expressions for single electron relaxation $\delta n_{k_{0}}(k)$ that will be given in a 
forthcoming publication.
 
This detailed analysis shows that, at zero temperature, $\delta n_{k_{0}}(k)$ has a quasiparticle peak at $k=k_0$ and a 
non-zero regular part for $-k_0 \leq k < k_0$ : 
$\delta n_{k_{0}}(k) = Z(k_0)\delta(k-k_0) + \delta n_{k_{0}}^r(k)$. 
Finally, the sum rule $Z(k_0) + \int_{-k_0}^{k_0}\delta n_{k_{0}}^{r}(k) dk = 1$ reflects particle conservation.  

Fig. \ref{fig:Relaxation1}-a presents the regular part $\delta n_{k_{0}}^r(k)$ as a function of $(k-k_0)l$ at fixed $k_0$, and Fig. \ref{fig:Relaxation1}-b shows the weight of the 
quasiparticle peak $Z(k_0)$ in term of $k_0l$ for $R=50\ \Omega$ and $l/v_{F}R_{K}C = 0.5$. 
In two limiting regimes, energy relaxation can be described using
a single particle relaxation approach by introducing an 
appropriate effective environment and, at low energy, taking into account the Pauli exclusion
principle.

Close enough to the Fermi surface  (see inset Fig. \ref{fig:Relaxation1}-a), single 
electron relaxation can be described using a simple relaxation model :
for $0\leq k < k_0$, $\delta n_{k_{0}}^r(k) \simeq p(k_0-k)$ where $p(q)$ can be 
interpreted as the probability to lose momentum $q$. The remaining quasiparticle weight
$Z(k_0)$ is then related to $p$ by $p(q) = -Z^{'}(q)$. Note that the Fermi sea 
remains spectator ($\delta n_{k_{0}}^r(k) \simeq 0$ for $k<0$), since 
at low energies $t(\omega)$ is close to one, thus limiting the 
creation of electron/hole pairs in the relaxation process. In this regime, energy relaxation is limited
by the Pauli exclusion principle. For $R\neq 0$ and at low energy, photon emission is the dominant relaxation 
mechanism and the inelastic scattering probability $1-Z(k_0)$, which only depends on the equivalent circuit parameters, scales as
$(\frac{R_{q}}{R_{K}}-\frac{1}{2})(k_{0}R_{K}C_{\mu}v_{F})^2$ at low momentum. Consequently, 
the quasi particle is well defined close to the Fermi surface.

At high energies, $\delta n_{k_{0}}(k)$ splits into two
distinct contributions (see \textit{e.g} Fig. \ref{fig:Relaxation1}-a, $k_0l = 8,\ 13$). The first
one appears in Fourier space
around the Fermi level and corresponds to the electron/hole pairs
created by the incident electron. The second one is localized in Fourier space close to
$k_{0}$ and corresponds to the relaxation of the incoming electron
due to photon emission into the transmission line as well as electron/hole
pair creation inside the edge channel. In real space, in the limit of very high incident energy, it shows up as an effective decoherence
coefficient $\mathcal{D}(x-y)$ acting on the incident wavepacket
propagated during the time interval $[0,t]$:
\begin{equation}\label{eq:single particle UV}
\varphi(x)\varphi^*(y)\mapsto \varphi(x-v_{F}t)\varphi^*(y-v_{F}t)\,\mathcal{D}(x-y)\,.
\end{equation}
The decoherence coefficient $\mathcal{D}$ is then a product
of the extrinsic contribution \eqref{eq:decoherence:extrinsic} and an
intrinsic contribution associated with the imprints left within the
Fermi sea by the different positions appearing in the single
electron incident wavepacket.  
Finally $\mathcal{D}$ is obtained by
substituting $|r(\omega)|^2$  by $2\Re(1-t(\omega))$ into
\eqref{eq:decoherence:extrinsic}. 
This regime is similar to the dynamical Coulomb blockade \cite{Ingold:1992-1}, with the Fourier transform of 
$\mathcal{D}$ playing the role of $P(E)$ (see dashed lines on 
Figs. \ref{fig:Relaxation1}-a and  \ref{fig:Relaxation2}-a for the corresponding energy relaxation).
The case of an RC-circuit with $R=R_K/2$ exhibits clear deviations from these
two limiting approaches in the considered frequency range 
(see Fig. \ref{fig:Relaxation2}-a) thus showing the need of our non-perturbative approach. This
arises from the strong plasmon scattering in the relevant frequency range at larger
$R/R_{K}$ (compare figs. \ref{fig:Relaxation1}-c and \ref{fig:Relaxation2}-c). Note also the faster decay of the quasiparticle peak (compare figs. \ref{fig:Relaxation1}-b and \ref{fig:Relaxation2}-b).

\paragraph*{Conclusion}
In this letter, we have presented a complete theory of coherent single electron excitation relaxation in integer quantum Hall
edge channels.
The scattering between edge plasmonic and environmental modes determines
both the finite frequency admittances and relaxation properties of a coherent single electron
excitation. The latter can thus be computed exactly from the finite frequency admittances.
This approach can be
used to study decoherence and relaxation of single electron excitations due to
interactions within a single edge channel or to interchannel coupling in $\nu=2$ quantum Hall systems.
It can also be generalized to fractional quantum Hall edges as well as to the case of
non chiral quantum wires. Finally, relaxation of non equilibrium distribution functions such as the one created by a biased quantum 
point contact requires taking into account all Keldysh correlators of currents \cite{Snyman:2008-1}.

\acknowledgements{We warmly thank Ch.~Glattli, B.~Pla\c{c}ais, F.~Pierre and P.~Roche for useful discussions and remarks.}


\begin{thebibliography}{24}
\expandafter\ifx\csname natexlab\endcsname\relax\def\natexlab#1{#1}\fi
\expandafter\ifx\csname bibnamefont\endcsname\relax
  \def\bibnamefont#1{#1}\fi
\expandafter\ifx\csname bibfnamefont\endcsname\relax
  \def\bibfnamefont#1{#1}\fi
\expandafter\ifx\csname citenamefont\endcsname\relax
  \def\citenamefont#1{#1}\fi
\expandafter\ifx\csname url\endcsname\relax
  \def\url#1{\texttt{#1}}\fi
\expandafter\ifx\csname urlprefix\endcsname\relax\def\urlprefix{URL }\fi
\providecommand{\bibinfo}[2]{#2}
\providecommand{\eprint}[2][]{\url{#2}}

\bibitem[{\citenamefont{Ji et~al.}(2003)\citenamefont{Ji, Chung, Sprinzak,
  Heiblum, Mahalu, and Shtrikman}}]{Ji:2003-1}
\bibinfo{author}{\bibfnamefont{Y.}~\bibnamefont{Ji}}
\bibnamefont{ {\it et al}},
  \bibinfo{journal}{Nature} \textbf{\bibinfo{volume}{422}},
  \bibinfo{pages}{415} (\bibinfo{year}{2003}).

\bibitem[{\citenamefont{Neder et~al.}(2007)\citenamefont{Neder, Ofek, Chung,
  Heiblum, Mahalu, and Umansky}}]{Neder:2007-2}
\bibinfo{author}{\bibfnamefont{I.}~\bibnamefont{Neder}}
\bibnamefont{ {\it et al}},
  \bibinfo{journal}{Nature} \textbf{\bibinfo{volume}{448}},
  \bibinfo{pages}{333} (\bibinfo{year}{2007}).

\bibitem[{\citenamefont{Roulleau
  et~al.}(2008{\natexlab{a}})\citenamefont{Roulleau, Portier, Glattli, Roche,
  Cavanna, Faini, Gennser, and Mailly}}]{Roulleau:2008-2}
\bibinfo{author}{\bibfnamefont{P.}~\bibnamefont{Roulleau}}
\bibnamefont{ {\it et al}},
  \bibinfo{journal}{Phys. Rev. Lett.} \textbf{\bibinfo{volume}{100}},
  \bibinfo{pages}{126802} (\bibinfo{year}{2008}{\natexlab{a}}).

\bibitem[{\citenamefont{F{\`e}ve et~al.}(2007)\citenamefont{F{\`e}ve, Mah{\'e},
  Berroir, Kontos, Pla{\c c}ais, Glattli, Cavanna, Etienne, and
  Jin}}]{Feve:2007-1}
\bibinfo{author}{\bibfnamefont{G.}~\bibnamefont{F{\`e}ve}}
\bibnamefont{ {\it et al}},
  \bibinfo{journal}{Science} \textbf{\bibinfo{volume}{316}},
  \bibinfo{pages}{1169} (\bibinfo{year}{2007}).

\bibitem[{\citenamefont{Hanbury-Brown and Twiss}(1956)}]{Hanbury:1956-1}
\bibinfo{author}{\bibfnamefont{R.}~\bibnamefont{Hanbury-Brown}}
  \bibnamefont{and} \bibinfo{author}{\bibfnamefont{R.}~\bibnamefont{Twiss}},
  \bibinfo{journal}{Nature} \textbf{\bibinfo{volume}{178}},
  \bibinfo{pages}{1046} (\bibinfo{year}{1956}).

\bibitem[{\citenamefont{Hong et~al.}(1987)\citenamefont{Hong, Ou, and
  Mandel}}]{Hong:1987-1}
\bibinfo{author}{\bibfnamefont{C.K.}~\bibnamefont{Hong}},
  \bibinfo{author}{\bibfnamefont{Z.Y.}~\bibnamefont{Ou}}, \bibnamefont{and}
  \bibinfo{author}{\bibfnamefont{L.}~\bibnamefont{Mandel}},
  \bibinfo{journal}{Phys. Rev. Lett.} \textbf{\bibinfo{volume}{59}},
  \bibinfo{pages}{2044} (\bibinfo{year}{1987}).

\bibitem[{\citenamefont{Ol'khovskaya et~al.}(2008)\citenamefont{Ol'khovskaya,
  Splettstoesser, Moskalets, and B{\"u}ttiker}}]{Olkhovskaya:2008-1}
\bibinfo{author}{\bibfnamefont{S.}~\bibnamefont{Ol'khovskaya}},
  \bibinfo{author}{\bibfnamefont{J.}~\bibnamefont{Splettstoesser}},
  \bibinfo{author}{\bibfnamefont{M.}~\bibnamefont{Moskalets}},
  \bibnamefont{and}
  \bibinfo{author}{\bibfnamefont{M.}~\bibnamefont{B{\"u}ttiker}},
  \bibinfo{journal}{Phys. Rev. Lett.} \textbf{\bibinfo{volume}{101}},
  \bibinfo{pages}{166802} (\bibinfo{year}{2008}).

\bibitem[{\citenamefont{Seelig and B{\"u}ttiker}(2001)}]{Seelig:2001-1}
\bibinfo{author}{\bibfnamefont{G.}~\bibnamefont{Seelig}} \bibnamefont{and}
  \bibinfo{author}{\bibfnamefont{M.}~\bibnamefont{B{\"u}ttiker}},
  \bibinfo{journal}{Phys. Rev. {\bf B}} \textbf{\bibinfo{volume}{64}},
  \bibinfo{pages}{245313} (\bibinfo{year}{2001}).

\bibitem[{\citenamefont{Chalker et~al.}(2007)\citenamefont{Chalker, Gefen, and
  Veillette}}]{Chalker:2007-1}
\bibinfo{author}{\bibfnamefont{J.T.}~\bibnamefont{Chalker}},
  \bibinfo{author}{\bibfnamefont{Y.}~\bibnamefont{Gefen}}, \bibnamefont{and}
  \bibinfo{author}{\bibfnamefont{M.Y.}~\bibnamefont{Veillette}},
  \bibinfo{journal}{Phys. Rev. {\bf B}} \textbf{\bibinfo{volume}{76}},
  \bibinfo{pages}{085320} (\bibinfo{year}{2007}).

\bibitem[{\citenamefont{Roulleau
  et~al.}(2008{\natexlab{b}})\citenamefont{Roulleau, Portier, Roche, Cavanna,
  Faini, Gennser, and Mailly}}]{Roulleau:2008-1}
\bibinfo{author}{\bibfnamefont{P.}~\bibnamefont{Roulleau}}
\bibnamefont{ {\it et al}},
  \bibinfo{journal}{Phys. Rev. Lett.} \textbf{\bibinfo{volume}{101}},
  \bibinfo{pages}{186803} (\bibinfo{year}{2008}{\natexlab{b}}).

\bibitem[{\citenamefont{Sukhorukov and Cheianov}(2007)}]{Sukhorukov:2007-2}
\bibinfo{author}{\bibfnamefont{E.V.}~\bibnamefont{Sukhorukov}} \bibnamefont{and}
  \bibinfo{author}{\bibfnamefont{V.V.}~\bibnamefont{Cheianov}},
  \bibinfo{journal}{Phys. Rev. Lett.} \textbf{\bibinfo{volume}{99}},
  \bibinfo{pages}{156801} (\bibinfo{year}{2007}).

\bibitem[{\citenamefont{Levkivskyi and Sukhorukov}(2008)}]{Levkivskyi:2008-1}
\bibinfo{author}{\bibfnamefont{I.P.}~\bibnamefont{Levkivskyi}} \bibnamefont{and}
  \bibinfo{author}{\bibfnamefont{E.V.}~\bibnamefont{Sukhorukov}},
  \bibinfo{journal}{Phys. Rev. {\bf B}} \textbf{\bibinfo{volume}{78}},
  \bibinfo{pages}{045322} (\bibinfo{year}{2008}).

\bibitem[{\citenamefont{Neuenhahn and Marquardt}(2009)}]{Neuenhahn:2009-1}
\bibinfo{author}{\bibfnamefont{C.}~\bibnamefont{Neuenhahn}} \bibnamefont{and}
  \bibinfo{author}{\bibfnamefont{F.}~\bibnamefont{Marquardt}},
  \bibinfo{journal}{Phys. Rev. Lett.} \textbf{\bibinfo{volume}{102}},
  \bibinfo{pages}{046806} (\bibinfo{year}{2009}).

\bibitem[{\citenamefont{Altimiras et~al.}(2009)\citenamefont{Altimiras,
  le~Sueur, Gennser, Cavanna, Mailly, and Pierre}}]{Altimiras:2009-1}
\bibinfo{author}{\bibfnamefont{C.}~\bibnamefont{Altimiras}}
\bibnamefont{ {\it et al}},
  (\bibinfo{year}{2009}), \bibinfo{note}{Nature Physics (in press)}.

\bibitem[{\citenamefont{Pines and Nozi{\`e}res}(1966)}]{Nozieres-Pines}
\bibinfo{author}{\bibfnamefont{D.}~\bibnamefont{Pines}} \bibnamefont{and}
  \bibinfo{author}{\bibfnamefont{P.}~\bibnamefont{Nozi{\`e}res}},
  \emph{\bibinfo{title}{The theory of quantum liquids}}
  (\bibinfo{publisher}{Perseus Book}, \bibinfo{year}{1966}).

\bibitem[{\citenamefont{Pr{\^e}tre et~al.}(1996)\citenamefont{Pr{\^e}tre,
  Thomas, and B{\"u}ttiker}}]{Pretre:1996-1}
\bibinfo{author}{\bibfnamefont{A.}~\bibnamefont{Pr{\^e}tre}},
  \bibinfo{author}{\bibfnamefont{H.}~\bibnamefont{Thomas}}, \bibnamefont{and}
  \bibinfo{author}{\bibfnamefont{M.}~\bibnamefont{B{\"u}ttiker}},
  \bibinfo{journal}{Phys. Rev. {\bf B}} \textbf{\bibinfo{volume}{54}},
  \bibinfo{pages}{8130} (\bibinfo{year}{1996}).

\bibitem[{\citenamefont{Yurke}(1984)}]{Yurke:1984-1}
\bibinfo{author}{\bibfnamefont{B.}~\bibnamefont{Yurke}}, \bibnamefont{and}
  \bibinfo{author}{\bibfnamefont{J.S.}~\bibnamefont{Denker}}, 
  \bibinfo{journal}{Phys. Rev. {\bf A}} \textbf{\bibinfo{volume}{29}},
  \bibinfo{pages}{1419} (\bibinfo{year}{1984}).

\bibitem[{\citenamefont{F{\`e}ve et~al.}(2008)\citenamefont{F{\`e}ve,
  Degiovanni, and Jolicoeur}}]{Degio25}
\bibinfo{author}{\bibfnamefont{G.}~\bibnamefont{F{\`e}ve}},
  \bibinfo{author}{\bibfnamefont{P.}~\bibnamefont{Degiovanni}},
  \bibnamefont{and}
  \bibinfo{author}{\bibfnamefont{T.}~\bibnamefont{Jolicoeur}},
  \bibinfo{journal}{Phys. Rev. {\bf B}} \textbf{\bibinfo{volume}{77}},
  \bibinfo{pages}{035308} (\bibinfo{year}{2008}).

\bibitem[{\citenamefont{Blanter et~al.}(1998)\citenamefont{Blanter, Hekking,
  and B{\"u}ttiker}}]{Blanter:1998-1}
\bibinfo{author}{\bibfnamefont{Y.M.}~\bibnamefont{Blanter}},
  \bibinfo{author}{\bibfnamefont{F.W.J.}~\bibnamefont{Hekking}}, \bibnamefont{and}
  \bibinfo{author}{\bibfnamefont{M.}~\bibnamefont{B{\"u}ttiker}},
  \bibinfo{journal}{Phys. Rev. Lett.} \textbf{\bibinfo{volume}{81}},
  \bibinfo{pages}{1925} (\bibinfo{year}{1998}).

\bibitem[{\citenamefont{Safi}(1999)}]{Safi:1999-1}
\bibinfo{author}{\bibfnamefont{I.}~\bibnamefont{Safi}}, \bibinfo{journal}{Eur.
  Phys. J. {\bf D}} \textbf{\bibinfo{volume}{12}}, \bibinfo{pages}{451}
  (\bibinfo{year}{1999}).

\bibitem[{\citenamefont{B{\"u}ttiker et~al.}(1993)\citenamefont{B{\"u}ttiker,
  Thomas, and Pr{\^e}tre}}]{Buttiker:1993-2}
\bibinfo{author}{\bibfnamefont{M.}~\bibnamefont{B{\"u}ttiker}},
  \bibinfo{author}{\bibfnamefont{H.}~\bibnamefont{Thomas}}, \bibnamefont{and}
  \bibinfo{author}{\bibfnamefont{A.}~\bibnamefont{Pr{\^e}tre}},
  \bibinfo{journal}{Phys. Lett. A} \textbf{\bibinfo{volume}{180}} \bibinfo{pages}{364}
  (\bibinfo{year}{1993}).

\bibitem[{\citenamefont{Gabelli et~al.}(2006)\citenamefont{Gabelli, F{\`e}ve,
  Berroir, Pla{\c c}ais, Cavanna, Etienne, Jin, and Glattli}}]{Gabelli:2006-1}
\bibinfo{author}{\bibfnamefont{J.}~\bibnamefont{Gabelli}}
\bibnamefont{ {\it et al}},
  \bibinfo{journal}{Science} \textbf{\bibinfo{volume}{313}},
  \bibinfo{pages}{499} (\bibinfo{year}{2006}).

\bibitem[{\citenamefont{Ingold and Nazarov}(1992)}]{Ingold:1992-1}
\bibinfo{author}{\bibfnamefont{G.-L.} \bibnamefont{Ingold}} \bibnamefont{and}
  \bibinfo{author}{\bibfnamefont{Y.}~\bibnamefont{Nazarov}},
  \emph{\bibinfo{title}{Single charge tunneling}} (\bibinfo{publisher}{Plenum
  Press, New York}, \bibinfo{year}{1992}), vol. \bibinfo{volume}{294} of
  \emph{\bibinfo{series}{NATO ASI Series B}}, chap. \bibinfo{chapter}{Charge
  tunneling rates in ultrasmall junctions}, pp. \bibinfo{pages}{21--107}.

\bibitem[{\citenamefont{Snyman and Nazarov}(2008)}]{Snyman:2008-1}
\bibinfo{author}{\bibfnamefont{I.}~\bibnamefont{Snyman}} \bibnamefont{and}
  \bibinfo{author}{\bibfnamefont{Y.V.}~\bibnamefont{Nazarov}},
  \bibinfo{journal}{Phys. Rev. {\bf B}} \textbf{\bibinfo{volume}{77}},
  \bibinfo{pages}{165118} (\bibinfo{year}{2008}).

\end{thebibliography}

\end{document}